\documentclass[12pt,aps,prl,preprint]{revtex4}
\usepackage{graphicx} \usepackage{subfigure}
\usepackage{stdclsdv} \usepackage{array} \usepackage{amsmath}

\begin{document}

\title{Multiple ordering transitions in a chiral liquid}

  \author{Pierre Ronceray} \affiliation{Laboratoire de
      Physique Th\'eorique et Mod\`eles Statistiques, Universit\'e
      Paris-Sud,
      B\^at. 100, 91405 Orsay Cedex, France\\
    D\'epartement de Physique, \'Ecole Normale Sup\'erieure,
      75005 Paris}

\author{and Peter Harrowell} \affiliation{School of Chemistry,
University of Sydney, Sydney N.S.W. 2006, Australia}

\begin{abstract}
  We present here a numerical study of a lattice model of a
  chiral liquid. The low symmetry of the favoured local structure
  depresses the freezing point to reveal an exotic liquid-liquid
  transition characterised by the appearance of an extended chirality,
  prior to freezing. What mechanisms impede crystallisation in liquids with low molecular
  symmetry ? The ordered liquid can be readily supercooled to
  zero temperature, as the combination of critical slowing down and
  competing crystal polymorphs results in a dramatically slow
  crystallisation process.
\end{abstract}


\maketitle

A tray of coins of the same size will, when gently shaken and
inclined, organize into a regular triangular lattice. This spontaneous
appearance of an extended structure can be rationalized in terms of
the high symmetry of the particles involved and the spatial
homogeneity that such symmetry implies. Replace the coins in the tray
with jigsaw puzzle pieces, even identical ones, and any expectation of
periodic ordering is greatly diminished. Why? The absence of a
sufficiently well packed crystal structure, the competition between
different local arrangements, the slow kinetics of rearrangement and
the increased entropy of the disordered state are all plausible
reasons for the failure of ordering of the puzzle pieces.  The very
number of possible explanations highlights our uncertainty about
collective ordering when low symmetry particles are
involved. Extensive theoretical studies of a variety of liquid crystal
phases - nematic~\cite{nematic}, cholesteric~\cite{chol},
discotic~\cite{discotic} and cubatic~\cite{cubic} - have provided deep
insight into the order associated with anisotropic particles that
retain a degree of rotational and reflection symmetry (or, in the case
of cholesterics, a perturbation from such symmetries). The kinetic
stability of molecular glass-forming liquids has long been loosely
attributed to some combination of low symmetry and conformational
flexibility.

 \begin{figure}[h]
  \includegraphics[width=0.21\textwidth]{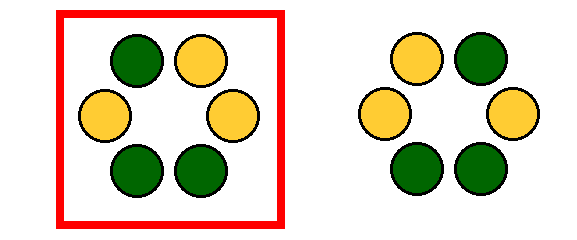}
  \caption{ \label{fig:structure} The pair of chiral local structures,
    with the favoured enantiomer boxed (up spins in yellow, down in
    green). Sites whose local spin environment correspond to this
    Favoured Local Structure and its rotational variants (regardless
    of the site's own spin) are given an energy of $-1$.  }
\end{figure}

\begin{figure}[h]
  \includegraphics[width=0.45\textwidth]{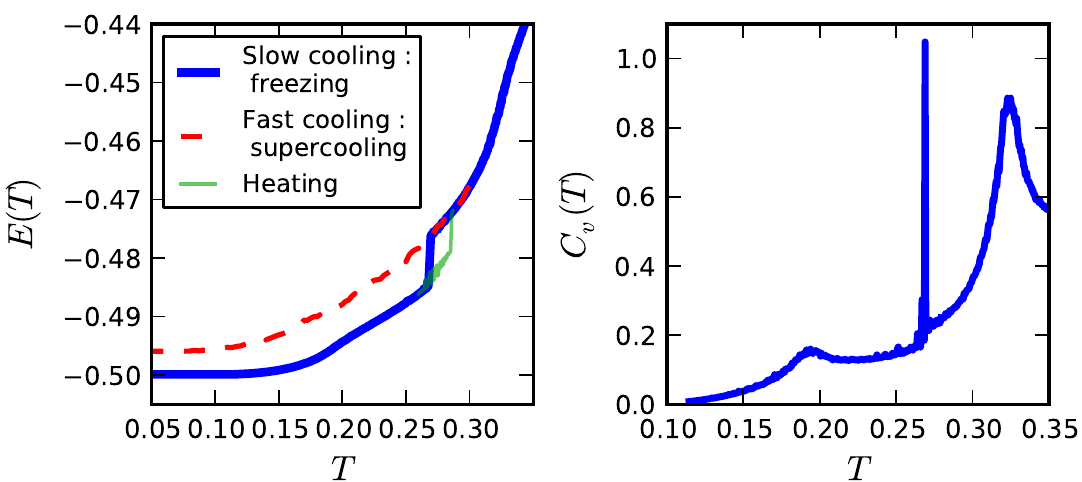}
  \caption{ \label{fig:energy} The temperature
    dependence of the average energy per site  and heat capacity
    (second panel). The high-temperature regime, which is not
    displayed here, is characteristic of liquids showing strong
    accumulation of local structures prior to
    crystallization, as discussed in \cite{ronceray2}.}
\end{figure}

We are interested in the behaviour of a liquid for which all geometric
symmetries are absent at the local level. The absence of reflection
symmetry means that the local structures are chiral. We consider a
simple lattice model with a restricted set of configurations, each
site being occupied by one of two objects (which we shall call spin up
or spin down).  For a given lattice we can enumerate every distinct
configuration of the nearest neighbours to a given site. We select the
favoured local structure by picking one such neighbourhood
configuration and assigning it an energy of -1, while all other local
arrangements are given an energy zero~\cite{ronceray1}. We consider
here the case of a 2D triangular lattice and its only possible pair of
chiral local structures (see inset in Figure~\ref{fig:energy}). A
discussion of the behaviour of the 3D face-centered cubic lattice and
its 39 pairs of chiral local structures will appear elsewhere.  We
have previously studied the groundstates~\cite{ronceray1}, liquid
entropy~\cite{ronceray2} and the freezing transitions~\cite{ronceray3}
in the 2D lattice model for the non-chiral local structures and shown
that freezing occurs with little or no evidence of supercooling for
all non-chiral favoured local structures. In this paper we report on
the dramatically different situation when the chiral local structure
is favoured.

\begin{figure*}[t] \centering
  \includegraphics[width=0.72\textwidth]{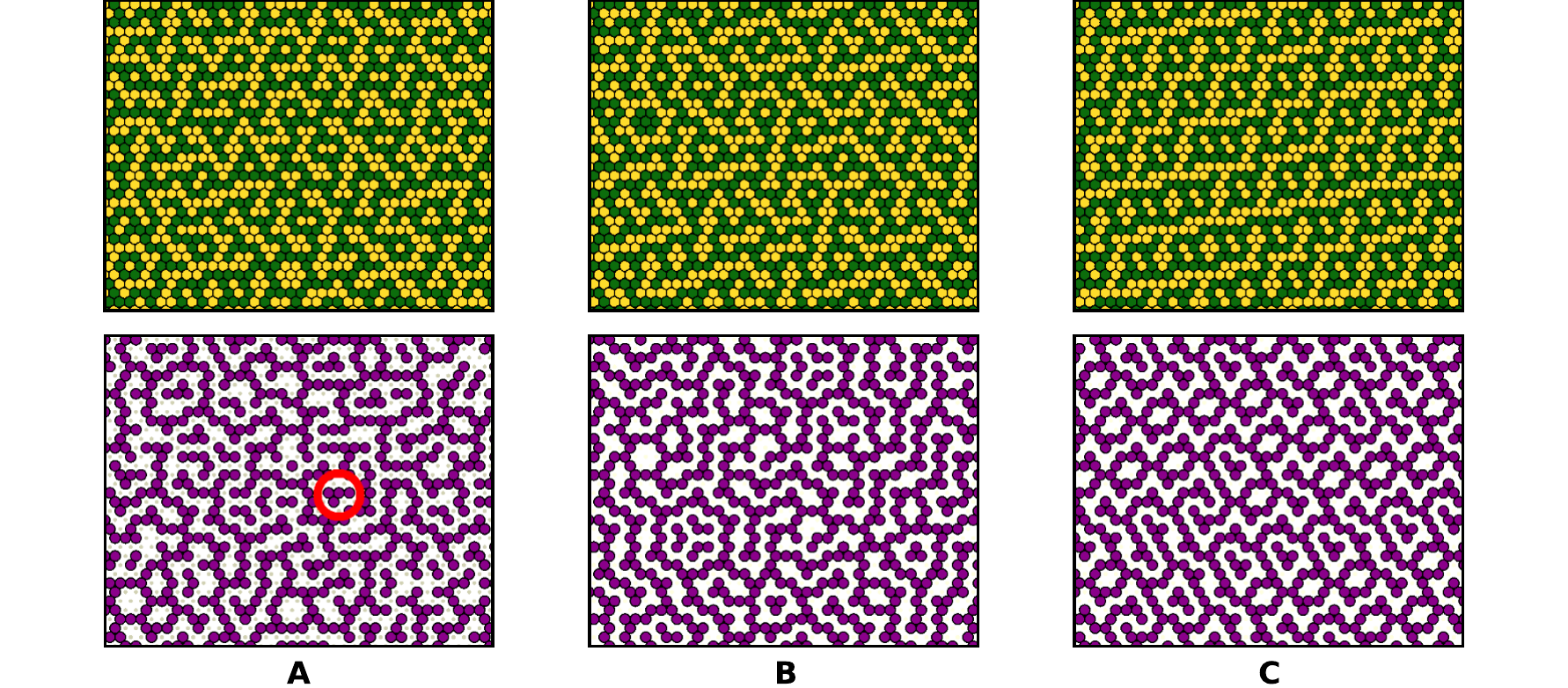}
  \caption{\label{fig:states} Some spin configurations of the
    system. The second row shows in purple the locus of sites that lie
    in the FLS. \textbf{A.} A liquid state just below the
    liquid-liquid transition. The liquid is highly structured, and has
    energy $-0.47$ - that is, $47\%$ of the lattice sites lie in the
    FLS. A chiral cluster of FLS, revealing the symmetry-breaking, is
    circled in red. \textbf{B.} A supercooled state at
    $T=0$. \textbf{C.} A ``crystalline'' state of energy $-1/2$. This
    state, while exhibiting long-range anisotropy, retains some freedom of configuration and extensive entropy,
    as it is a mixture of crystalline domains with no grain boundary
    energy cost. }
\end{figure*}

In Figure~\ref{fig:energy} we plot the average energy $E$ at
equilibrium (thick blue line) and the heat capacity $C_{V} = \partial
E/\partial T$ \emph{versus} temperature for heating and cooling
runs. We find clear evidence for two transitions in our model
liquid. A second-order transition occurs at $T_c \approx 0.325$ as
indicated by the peak in $C_V$. It separates the high-temperature
disordered state from an ordered liquid with an extended chirality, as
illustrated in Figure~\ref{fig:states} \textbf{A}. In the case of very
slow cooling rates, it is followed by a first order phase transition
at $0.26 \le T_f \le 0.29$ associated with significant hysteresis. The
first order transition corresponds to the liquid freezing into the
polycrystalline ordered state depicted in Figure~\ref{fig:states}
\textbf{C}. At faster cooling rates dashed line in
Figure~\ref{fig:energy}), the system gets stuck in a low-energy
supercooled state, as depicted in Figure~\ref{fig:states} \textbf{B}.

\begin{figure*}
  \centering
  \includegraphics[width=1.\textwidth]{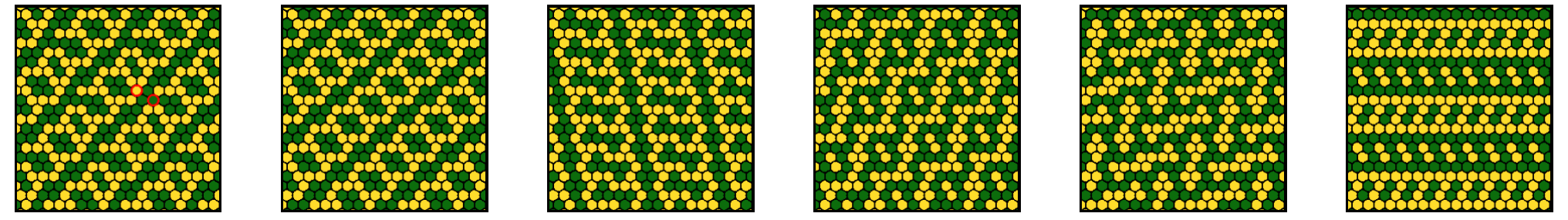}
  \caption{\label{fig:crystals} Some identified ideal crystalline
    structures for the chiral system, which illustrate its structural
    complexity. All these structures have an energy of $-1/2$ per
    site, and they are all anisotropic.  Note that in the leftmost one,
    the two sites outlined in red can be flipped simultaneously at no
    energy cost. This implies an extensive entropy for the ground
    state of the system. Only the rightmost crystalline structure preserves
    the $\Sigma$ symmetry. }
\end{figure*}

The ground state energy of the chiral system is $E_o = -1/2$, which
means that at most half of the lattice sites can be simultaneously in
the favoured local structure. This energy is considerably higher than
that of the racemic mixture, for which $E_o =
-4/5$~\cite{ronceray3}. The significant stability of the racemic
crystal over that of the chiral structure is probably not a chance
outcome. Over $90\%$ of chiral organic molecules crystallize
preferentially as racemic crystals consisting of equal numbers of both
enantiomers~\cite{crystal}. We have identified several crystal
polymorphs with an energy $-1/2$, six of which are depicted in
Figure~\ref{fig:crystals}. We have, however, only ever observed the
polycrystalline mixture shown in Figure~\ref{fig:states}, which has an
energy of exactly $-1/2$.

\begin{figure}[h]
  \includegraphics[width=0.22\textwidth]{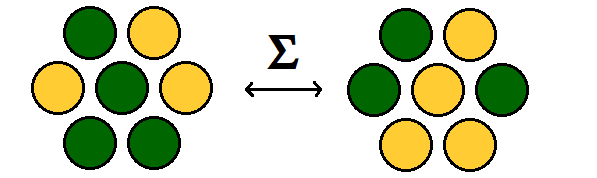}
  \caption{ \label{fig:symmetry} The only symmetry of the FLS, which
    we call $\Sigma$, consists in an inversion of all spins and a
    reflection.}
\end{figure}

\begin{figure}[h]
 \includegraphics[width=0.45\textwidth]{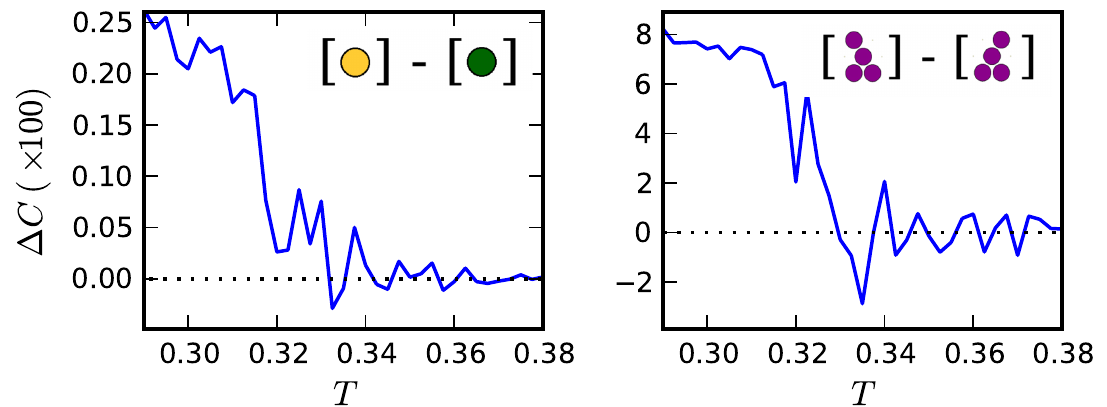}
 \caption{ \label{fig:order_parameter} \textbf{Left panel:} The magnetization,
   defined as the difference of concentrations $\Delta C$ of up and
   down spins, as a function of temperature. The critical point $T_c$
   is identified as the temperature below which it bifurcates to a
   non-zero value. \textbf{Right panel:} The extended chirality, defined as the
   difference of concentration of the two clusters of FLS represented,
   exhibits the same behaviour.}
\end{figure}

The freezing temperature $T_f$ can be related to the change in energy
$\Delta E$ and entropy $\Delta S$ on freezing by $T_{f}=\Delta
E/\Delta S$. The central point here is that any geometric
``awkwardness'' that produces a high energy crystal will typically
result in the depression of the freezing point. One consequence of the
low freezing point is that, in extending the temperature range of the
liquid, a greater accumulation of local structure will take place
prior to crystallization. How much local structure a liquid can
accumulate and how this accumulation ends are questions that have
arisen in the study of supercooled liquids. Suggested endpoints for a
supercooled liquid, as the fraction of favoured local structure
increases, are either an instability with respect to crystallization
or the formation of a glass. In our chiral lattice liquid we observe a
third outcome, a 2nd order liquid-liquid transition that occurs at
$T_c \approx 0.325$. Unlike the zero-field Ising model~\cite{ising},
our chiral Hamiltonian is {\it not} symmetric with respect to spin
inversion since the inversion of all spins will transform one
enantiomer into the other, a transformation which will involve a
change in energy as only one enantiomer is favoured. The only
elementary symmetry for our chiral Hamiltonian is the combination of a
global spin inversion with a reflection through a line (we shall refer
to this composite operation as $\Sigma$) as shown in
Figure~\ref{fig:symmetry}. This symmetry is spontaneously broken at
the second-order phase transition. We identified the driving force for
this symmetry breaking as the short-range entropic repulsion between
favoured local structures with different values for the central spin,
using the formalism introduced in~\cite{ronceray2}. Applying the
operation $\Sigma$ we recover a configuration unchanged in energy but
with a change in magnetization. It follows that magnetization
(\emph{i.e.} the difference between up and down spin concentrations)
represents an order parameter for the 2nd order phase transition
observed in the chiral liquid.  In Figure~\ref{fig:order_parameter} we
plot the temperature dependence of the magnetization, which bifurcates
as the temperature drops below $T_c$.

Because of the $\Sigma$ symmetry, the energy landscape -- that is, the
locus of the favoured local structures, as depicted in purple in
Figure~\ref{fig:states} -- is not chiral at high temperature. The two
states of different magnetization below $T_c$ correspond to the two
possible choices of ``handedness'' for the extended clusters of
overlapping favoured local structures, such as the cluster circled in
red in Figure~\ref{fig:states} \textbf{A} - so that the onset of a
global chirality in the liquid. This structural chirality can be
measured by plotting the difference of density between this extended
cluster and its enantiomer, as shown in the right panel on
Figure~\ref{fig:order_parameter}.  We categorize this transition as a
liquid-liquid transition since translational disorder is retained in
the low temperature phase along with a first order freezing
transition. Despite the presence of a global chirality, this is not a
liquid crystal in the usual sense because no global rotational
anisotropy is observed. It is well established~\cite{chemla} in
crystals that the optical activity (an observable aspect of chirality)
includes contributions from the global structure as well as from the
purely local chirality. $\alpha$-quartz, for example, is chiral
despite the fact that the elementary unit - the silica tetrahedron -
is achiral. The low temperature liquid phase described here is, to our
knowledge, the first example of a \emph{liquid} whose chirality
includes a contribution from the global structure.

\begin{figure}[ht]
  \includegraphics[width=0.3\textwidth]{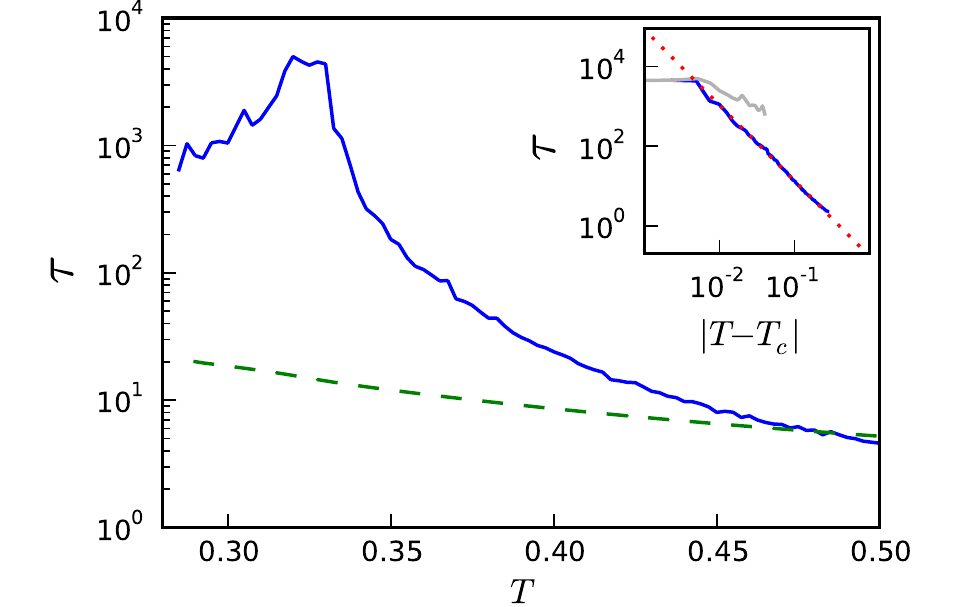}
  \caption{ \label{fig:time} Solid line: the characteristic
    autocorrelation time of the energy of the system, measured in
    Monte-Carlo Metropolis steps per site. It diverges at the
    second-order phase transition, revealing critical
    slowing-down. Dashed line: The spin persistence time, defined as
    the average time before a site flips. This quantity is not
    affected by the transition at $T_c$. \\ \textbf{Inset: } The autocorrelation
    time in log-log scale. Dashed line: power-law fit $ \tau = 0.17\
    |T-T_c|^{-1.9}$ with $T_c = 0.325$. The agreement is excellent at
    $T>T_c$. The behaviour at $T<T_c$ (light grey) remains
    unexplained.}
\end{figure}

The accumulation of local favoured structures has, in supercooled
liquids, been invoked as the origin of the substantial slowing down
observed in these systems on cooling~\cite{supercooled}. In the case
of our chiral lattice liquid the temperature dependence of {\it some}
relaxation times corresponds to the standard critical slowing
down. The relaxation time $\tau_{E}$ obtained from the autocorrelation
function of the energy fluctuations exhibits a divergence at the
transition temperature, as shown in Figure~\ref{fig:time}. At $T>T_c$,
we find that the temperature dependence of $\tau_{E}$ when approaching
the transition temperature $T_c$ is well described by a power law
$\tau_{E} \propto (T-T_c)^{-\gamma}$ where $\gamma \approx 1.9$, a
value close to exponent $2$ found for the critical slowing down in the
Ising model on the 2D square lattice~\cite{ising}. The decrease in
$\tau_{E}$ as the temperature is lowered below $T_c$ reflects the
decreasing amplitude of the energy fluctuations. Not all time scales
in the liquid are slaved to this critical slowing down. The
persistence time, defined as the average lifetime of the individual
spin states, shows no such singular temperature dependence (dashed
line in Figure~\ref{fig:time}). This is expected for a liquid-liquid
transition: the spin structures are transient and ``flow''.

Another time scale of physical significance is that of
crystallization. The first-order phase transition at $T\approx 0.26$
is observed only for cooling rates slower than $10^{8}$ Monte-Carlo
Metropolis steps per site and unit of temperature (for comparison, a
cooling rate of only $10^4$ leads to crystallisation in achiral
liquids in this model~\cite{ronceray1}.) This tremendously slow
process could be observed using a rejection-free Monte-Carlo
algorithm~\cite{bortz}. At faster cooling rate, the system gets
``stuck'' in one of many supercooled metastable states. Note that the
observed ground state, shown in Figure~\ref{fig:states}, shares the
same broken symmetry as the liquid state below $T_c$. (This broken
symmetry is shared by all crystalline groundstates shown on
Figure~\ref{fig:crystals}, except the rightmost one, which is never
observed.) This means that the crystallization rate will reflect the
slow relaxation time of the structural fluctuations of the chiral
ordered liquid state.  As shown in Figure~\ref{fig:energy}, there is
substantial hysteresis in the $E(T)$ curves about the freezing
transition. When a sufficiently slow heating rate is employed, melting
is observed at a temperature well below $T_c$, establishing the
separation of the two transitions.

The consequences of the low symmetry of the local stable structure can
be summarised as follows. The crystal state has high energy, a large
unit cell and many polymorphs. Because of the high energy of the
crystal, the stability of the liquid is extended to low temperatures
so that it accumulates a considerable amount of locally favoured
structures. This high density of FLS ultimately drives the
liquid-liquid transition reported here, associated with the appearance
of an extended chirality of the liquid structure itself. The
non-Arrhenius temperature dependence of relaxation dynamics of the
liquid can be directly associated with the slowing down associated
with this critical point. Crystallization from the ordered liquid is
extremely slow, so that supercooled liquid states are readily
observed, with an energy that closely approaches that of the crystal.

There is a considerable amount of literature on the build up of local
structures in liquids as they are supercooled and the possible
consequences of this ordering~\cite{supercooled}. A number of liquids
exhibit a liquid-liquid transitions in either at equilibrium
(sulfur~\cite{sulfur}, phosphorous~\cite{phosphorous} and
cerium~\cite{cerium}) or in the supercooled state
(silica~\cite{silica}, water~\cite{water}, silicon~\cite{silicon} and
triphenyl phosphite~\cite{TPP}). In a number of cases the
liquid-liquid transition appears to coincide with either the
instability of the metastable state with respect to
crystallization~\cite{molinero} or the glass
transition\cite{tanaka}. In the case of a number of models of atomic
mixtures, an accumulation of local order is observed without any sign
of a liquid-liquid transition~\cite{supercooled}. The low temperature
fate of such liquids remains unclear - the possibilities include an
instability with respect to crystallization (the Kauzmann solution),
the arrest into a glass state or some order-disorder transition not
yet observed. The chiral FLS model presented here provides a case
study of a ``complex'' liquid in which these unknowns are explicitly
resolved in the manner described in the preceding paragraph. The key
feature of the chiral FLS model is the separability of the
liquid-liquid transition and crystallization, the two associated
broken symmetries being sufficiently decoupled to permit the
occurrence of the two distinct transitions. Which aspects of a
Hamiltonian determine whether this decoupling occurs or not is, we suggest, an
important question for future research into the low temperature fate
of supercooled liquids.

{\bf Acknowledgements}\\ The authors thank Toby Hudson for valuable
discussions and Hanna Gr\"onqvist for careful proofreading. We
acknowledge funding from the \'{E}cole Normale Sup\'{e}rieure and the
Australian Research Council.

\end{document}